# Impact of Pointing Errors on the Performance of Mixed RF/FSO Dual-Hop Transmission Systems

Imran Shafique Ansari, *Student Member, IEEE*, Ferkan Yilmaz, *Member, IEEE*, and Mohamed-Slim Alouini, *Fellow, IEEE*

*Abstract*—In this work, the performance analysis of a dual-hop relay transmission system composed of asymmetric radio-frequency (RF)/free-space optical (FSO) links with pointing errors is presented. More specifically, we build on the system model presented in [1] to derive new exact closed-form expressions for the cumulative distribution function, probability density function, moment generating function, and moments of the end-to-end signal-to-noise ratio in terms of the Meijer's G function. We then capitalize on these results to offer new exact closed-form expressions for the higher-order amount of fading, average error rate for binary and $M$-ary modulation schemes, and the ergodic capacity, all in terms of Meijer's G functions. Our new analytical results were also verified via computer-based Monte-Carlo simulation results.

## I. INTRODUCTION

IN recent times, free-space optical (FSO) or optical wireless communication systems have gained an increasing interest due to its various characteristics including higher bandwidth and higher capacity compared to the traditional radio frequency (RF) communication systems. In addition, FSO links are license-free and hence are cost-effective relative to the traditional RF links. These features of FSO communication systems potentially enable solving the issues that the RF communication systems face due to the expensive and scarce spectrum [2]–[6]. However, the atmospheric turbulence may lead to a significant degradation in the performance of the FSO communication systems [2]. Additionally, thermal expansion, dynamic wind loads, and weak earthquakes result in the building sway phenomenon that causes vibration of the transmitter beam leading to a misalignment between transmitter and receiver known as pointing error. These pointing errors may lead to significant performance degradation and are a serious issue in urban areas, where the FSO equipments are placed on high-rise buildings [7]–[9].

On the other hand, relaying technology has gained enormous attention for quite a while now since it not only provides wider and energy-efficient coverage but also increased capacity in the wireless communication systems. As such many efforts have been made to study the relay system performance under various fading conditions [10]–[13]. These independent studies consider symmetric channel conditions i.e. the links at the hops are similar in terms of the fading distributions though it is more practical to experience different/asymmetric link conditions at different hops i.e. each link may differ in the channel conditions from the other link [1], [14]–[17]. This is due to the fact that the signals on each hop are transmitted either via different communications systems or the signals might have to commute through physically different paths. For instance, as proposed in [1], a relaying system based on both FSO as well as RF characteristics can be expected to be more adaptive and constitute an effective communication system in a real-life environment.

The model utilized in our work is similar to the one presented in [1]. Although [1] lacks the motivation behind such a model, we understand and proceed with such a model based on the following explanation. Considering an uplink scenario, besides all the advantages of FSO over RF, that very much motivates this work is the concept of multiplexing i.e. we can multiplex users with RF only capability into a single FSO link. This comes with the reasoning that there exists a connectivity gap between the backbone network and the last-mile access network and hence this last mile connectivity can be delivered via high-speed FSO links [18]. For instance, in developing countries where there might not be much of a fiber optic structure and hence to increase its reach and bandwidth to the last mile, it will require huge amount of economic resources to dig up the current brown-field. It will be much better to simply install FSO transmitters and detectors on the high-rise buildings and cover the last mile by having the users with RF capability to communicate via their respective RF bands and let the rest be taken care of by the FSO links to get it through to the backbone as can be observed from Fig. 1. This multiplexing feature will avoid the bottleneck situation for the system capacity and in fact be a faster option relative to traditional RF-RF communications wherein multiple RF's being sent through a single FSO at once. Hence, there are two outstanding features, among others, of this system to make it very advantageous over the current traditional system. Firstly, maximum possible RF messages can be aggregated into a single FSO link thereby utilizing the system to the maximum possible capacity. Simultaneously, the system benefits from another feature of having the RF link always available irrespective of FSO transmission since the RF and the FSO operate on completely different sets of frequencies allowing for no interference between them at any instant. Therefore, the RF frequency bands can be utilized by other possible devices/users around in range to their benefit while the FSO link is yet under operation. Above all, having FSO will avoid any sort of interference(s) also due to its point-to-point transmission feature unlike RF where the transmission is a broadcast leading to possible interference(s). In Fig. 1, there exists no fiber optics structure between the

Imran Shafique Ansari, Ferkan Yilmaz, and Mohamed-Slim Alouini are with the Computer, Electrical, and Mathematical Sciences, and Engineering (CEMSE) Division at King Abdullah University of Science and Technology (KAUST), Thuwal, Makkah Province, Saudi Arabia (e-mail: {imran.ansari, ferkan.yilmaz, slim.alouini}@kaust.edu.sa).

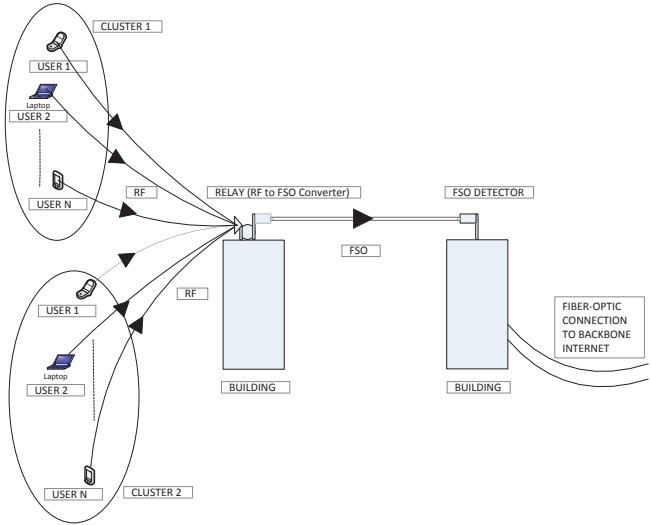

Fig. 1. System model block diagram of an asymmetric mixed RF/FSO dual-hop transmission system.

buildings. Since similar optical transmitters and detectors are used for FSO and fiber optics, similar bandwidth capabilities are achievable [18]. Therefore, this will get the required job done saving numerous amount of economic resources by utilizing FSO instead of digging up the current brownfield to install fiber optics between the different buildings. Another set of motivation behind such a system involves a fact that the users are mostly mobile and with only RF capabilities (no FSO capabilities). Installing FSO capability on these mobile users does not seem to be a justified approach. Simultaneously, we also fall short of bandwidth (BW) every now and then. Hence, to save on BW and to save on the economic resources by avoiding unnecessary modifications to the current mobile devices, we have introduced such a system wherein the users remain as is with RF only capability(s) and yet can be part of and/or make use of the FSO featured network. In another instance, we can think of a building floor (femto-cell in a heterogeneous network) where the users can send and receive through the backbone via FSO transmitter and detector respectively, placed at one of the corners of that floor. This FSO transmitter/detector can communicate with other such devices over other high-rise buildings and ultimately hop to the backbone. Additionally, to increase the spectral efficiency of such a system, we can study the effects on the performance of the system by selection of $N$-best users to be multiplexed. Hence, to perform such a study and due to space limitations, this manuscript tackles only the simplest possible scenario/special case of such a heterogeneous system and studies its statistical characteristics and ultimately the performance measures with on-going work on ultimately addressing all the issues mentioned above and/or as shown in Fig. 1 and beyond.

However, the results presented in [1] were derived under the assumption of non-pointing errors in the FSO link but were limited to cumulative distribution function (CDF)/outage probability (OP). In this work, we build on the model presented in [1] to study the impact of pointing errors on the performance of asymmetric RF/FSO dal-hop transmission systems with fixed gain relays. For instance, we derive the CDF, probability density function (PDF), moment generating function (MGF), and moments of the end-to-end signal-to-noise ratio (SNR) of such systems. We then apply this statistical characterization of the SNR to derive closed-form expressions of the higher-order amount of fading (AF), average bit-error rate (BER) of binary modulation schemes, average symbol error rate (SER) of $M$-ary amplitude modulation (M-AM), $M$-ary phase shift keying (M-PSK) and $M$-ary quadrature amplitude modulation (M-QAM), and the ergodic capacity in terms of Meijer's G functions.

## II. CHANNEL AND SYSTEM MODELS

We employ the same model as was employed in [1] and hence, the end-to-end SNR can be given as $\gamma = \frac{\gamma_1 \gamma_2}{\gamma_2 + C}$, where $\gamma_1$ represents the SNR of the RF hop i.e. S-R link, $\gamma_2$ represents the SNR of the FSO hop i.e. R-D link, and $C$ is a fixed relay gain [1], [10], [19].

The RF link (i.e. S-R link) is assumed to follow Rayleigh fading whose SNR follows an exponential distribution, parameterized by the average SNR $\overline{\gamma}_1$ of the S-R link, with a PDF given by $f_{\gamma_1}(\gamma_1) = 1/\overline{\gamma}_1 \exp(-\gamma_1/\overline{\gamma}_1)$ [19]. On the other hand, it is assumed that the FSO link (i.e. R-D link) experiences Gamma-Gamma fading with pointing error impairments whose SNR PDF is given under indirect modulation/direct detection (IM/DD) by [8, Eq. (12)], [9, Eq. (20)] that can be expressed in a simpler form by utilizing [20, Eq. (6.2.4)], as

$$f_{\gamma_2}(\gamma_2) = \frac{\xi^2}{2\gamma_2\,\Gamma(\alpha)\Gamma(\beta)}\, \mathrm{G}_{1,3}^{3,0}\!\left[\alpha\beta\sqrt{\frac{\gamma_2}{\overline{\gamma}_2}} \,\bigg|\, \begin{matrix} \xi^2+1 \\ \xi^2, \alpha, \beta \end{matrix}\right], \quad (1)$$

where $\overline{\gamma}_2$ is the average SNR of the R-D link, $\alpha$ and $\beta$ are the fading parameters related to the atmospheric turbulence conditions [3]–[5] with lower values of $\alpha$ and $\beta$ indicating severe atmospheric turbulence conditions, $\xi$ is the ratio between the equivalent beam radius at the receiver and the pointing error displacement standard deviation (jitter) at the receiver [8], [9], $\Gamma(.)$ is the Gamma function as defined in [21, Eq. (8.310)], and $\mathrm{G}(.)$ is the Meijer's G function as defined in [21, Eq. (9.301)].

## III. CLOSED-FORM STATISTICAL CHARACTERISTICS

### A. Cumulative Distribution Function

The CDF is given by [10]

$$F_\gamma \gamma = \Pr\left[\frac{\gamma_1\gamma_2}{\gamma_2 + C} < \gamma\right], \quad (2)$$

which can be written as

$$F_\gamma(\gamma) = \int_0^\infty \Pr\left[\frac{\gamma_1 \gamma_2}{\gamma_2 + C} < \gamma | \gamma_2\right] f_{\gamma_2}(\gamma_2) \, d\gamma_2$$
$$= 1 - \frac{\alpha \beta \xi^2}{2\sqrt{\overline{\gamma_2}} \, \Gamma(\alpha) \Gamma(\beta)} \exp(-\gamma/\overline{\gamma}_1) \quad (3)$$
$$\times \int_0^\infty (1/\sqrt{\gamma_2}) \exp(-\gamma C/(\gamma_2 \overline{\gamma}_1))$$
$$\times G_{1,3}^{3,0}\left[\alpha \beta \sqrt{\frac{\gamma_2}{\overline{\gamma}_2}} \, \bigg| \, \begin{matrix} \xi^2 \\ \xi^2-1, \alpha-1, \beta-1 \end{matrix}\right] d\gamma_2,$$

Using [22, Eq. (07.34.03.0228.01)], we can rewrite $\exp(-\gamma C/(\gamma_2 \overline{\gamma}_1))$ as $G_{0,1}^{1,0}\left[\frac{\gamma C}{\gamma_2 \overline{\gamma}_1} \, \big| \, 0\right]$. Further using [20, Eq. (6.2.2)], $G_{0,1}^{1,0}\left[\frac{\gamma C}{\gamma_2 \overline{\gamma}_1} \, \big| \, 0\right]$ can be alternated to $G_{1,0}^{0,1}\left[\frac{\gamma_2 \overline{\gamma}_1}{\gamma C} \, \big| \, 1\right]$. Now, along with the above modifications, applying [23, Eq. (21)] to (3), and some simple algebraic manipulations along with utilizing [20, Eq. (6.2.4)], the CDF of $\gamma$ can be shown to be given by

$$F_\gamma(\gamma) = 1 - A_1 \exp(-\gamma/\overline{\gamma}_1) \, G_{1,6}^{6,0}\left[\frac{B}{\overline{\gamma}_1} \gamma \, \bigg| \, \begin{matrix} \kappa_1 \\ \kappa_2 \end{matrix}\right], \quad (4)$$

where $\kappa_1 = \frac{\xi^2}{2} + 1$, $\kappa_2 = \frac{\xi^2}{2}, \frac{\alpha}{2}, \frac{\alpha+1}{2}, \frac{\beta}{2}, \frac{\beta+1}{2}, 0$, $A_1 = \frac{\xi^2 \, 2^{\alpha+\beta}}{8\pi \Gamma(\alpha)\Gamma(\beta)}$, and $B = \frac{(\alpha\beta)^2 C}{16 \overline{\gamma}_2}$. For the non-pointing errors case, when $\xi \to \infty$, it can be easily shown that the CDF in (4) converges to

$$F_\gamma(\gamma) = 1 - A_2 \exp(-\gamma/\overline{\gamma}_1) \, G_{0,5}^{5,0}\left[\frac{B}{\overline{\gamma}_1} \gamma \, \bigg| \, \begin{matrix} - \\ \kappa_3 \end{matrix}\right], \quad (5)$$

where $\kappa_3 = \frac{\alpha}{2}, \frac{\alpha+1}{2}, \frac{\beta}{2}, \frac{\beta+1}{2}, 0$, and $A_2 = \frac{2^{\alpha+\beta}}{4\pi \Gamma(\alpha)\Gamma(\beta)}$ in agreement with [1, Eq. (15)].

### B. Probability Density Function

Differentiating (4) with respect to $\gamma$, using the product rule then utilizing [22, Eq. (07.34.20.0001.01)], we obtain after some algebraic manipulations the PDF in exact closed-form in terms of Meijer's G functions as

$$f_\gamma(\gamma) = \frac{A_1}{2\gamma \overline{\gamma}_1} \exp(-\gamma/\overline{\gamma}_1) \left(2\overline{\gamma}_1 \, G_{0,5}^{5,0}\left[\frac{B}{\overline{\gamma}_1} \gamma \, \bigg| \, \begin{matrix} - \\ \kappa_3 \end{matrix}\right] \right.$$
$$\left. + (2\gamma - \xi^2 \overline{\gamma}_1) \, G_{1,6}^{6,0}\left[\frac{B}{\overline{\gamma}_1} \gamma \, \bigg| \, \begin{matrix} \kappa_1 \\ \kappa_2 \end{matrix}\right]\right). \quad (6)$$

For the non-pointing errors case, when $\xi \to \infty$, it can be easily shown that the PDF in (6) converges to

$$f_\gamma(\gamma) = \frac{A_2}{16 \overline{\gamma}_1 \overline{\gamma}_2} \exp(-\gamma/\overline{\gamma}_1) \left(16 \overline{\gamma}_2 \, G_{0,5}^{5,0}\left[\frac{B}{\overline{\gamma}_1} \gamma \, \bigg| \, \begin{matrix} - \\ \kappa_3 \end{matrix}\right] \right.$$
$$\left. + (\alpha\beta)^2 C \, G_{0,5}^{5,0}\left[\frac{B}{\overline{\gamma}_1} \gamma \, \bigg| \, \begin{matrix} - \\ \kappa_4 \end{matrix}\right]\right), \quad (7)$$

where $\kappa_4 = \frac{\alpha}{2} - 1, \frac{\alpha-1}{2}, \frac{\beta}{2} - 1, \frac{\beta-1}{2}, 0$.

### C. Moment Generating Function

The MGF defined as $\mathcal{M}_\gamma(s) \triangleq \mathbb{E}[e^{-\gamma s}]$ can be expressed in terms of CDF as $\mathcal{M}_\gamma(s) = s \int_0^\infty e^{-\gamma s} F_\gamma(\gamma) d\gamma$. Using this equation by placing (4) into it and utilizing [21, Eq. (7.813.1)], we get after some manipulations the MGF of $\gamma$ as

$$\mathcal{M}_\gamma(s) = 1 - \frac{s \, A_1}{s + 1/\overline{\gamma}_1} \, G_{2,6}^{6,1}\left[\frac{B}{s \overline{\gamma}_1 + 1} \, \bigg| \, \begin{matrix} 0, \kappa_1 \\ \kappa_2 \end{matrix}\right]. \quad (8)$$

When $\xi \to \infty$ (i.e. non-pointing error case), the MGF in (8) can be easily shown to converge to

$$\mathcal{M}_\gamma(s) = 1 - \frac{s \, A_2}{s + 1/\overline{\gamma}_1} \, G_{1,5}^{5,1}\left[\frac{B}{s \overline{\gamma}_1 + 1} \, \bigg| \, \begin{matrix} 0 \\ \kappa_3 \end{matrix}\right]. \quad (9)$$

### D. Moments

The moments defined as $\mathbb{E}[\gamma^n]$ can be expressed in terms of the complementary CDF (CCDF) $F_\gamma^c(\gamma) = 1 - F_\gamma(\gamma)$ as $\mathbb{E}[\gamma^n] = n \int_0^\infty \gamma^{n-1} F_\gamma^c(\gamma) d\gamma$. Now, using this equation by placing (4) into it and utilizing [21, Eq. (7.813.1)], we get the moments as

$$\mathbb{E}[\gamma^n] = n \, A_1 \, \overline{\gamma}_1^n \, G_{2,6}^{6,1}\left[B \, \bigg| \, \begin{matrix} 1-n, \kappa_1 \\ \kappa_2 \end{matrix}\right]. \quad (10)$$

When $\xi \to \infty$, the moments in (10) can be easily shown to converge to

$$\mathbb{E}[\gamma^n] = n \, A_2 \, \overline{\gamma}_1^n \, G_{1,5}^{5,1}\left[B \, \bigg| \, \begin{matrix} 1-n \\ \kappa_3 \end{matrix}\right]. \quad (11)$$

## IV. APPLICATIONS TO THE PERFORMANCE OF ASYMMETRIC RF/FSO RELAY TRANSMISSION SYSTEMS

### A. Higher-Order Amount of Fading

The AF is an important measure for the performance of a wireless communication system as it can be utilized to parameterize the distribution of the SNR of the received signal. In particular, the $n^{th}$-order AF for the instantaneous SNR $\gamma$ is defined as $AF_\gamma^{(n)} = \mathbb{E}[\gamma^n]/\mathbb{E}[\gamma]^n - 1$ [24]. Now, utilizing this equation by substituting (10) into it, we get the $n^{th}$-order AF as

$$AF_\gamma^{(n)} = n \, A_1^{1-n} \, G_{2,6}^{6,1}\left[B \, \bigg| \, \begin{matrix} 1-n, \kappa_1 \\ \kappa_2 \end{matrix}\right] G_{2,6}^{6,1}\left[B \, \bigg| \, \begin{matrix} 0, \kappa_1 \\ \kappa_2 \end{matrix}\right]^{-n} - 1. \quad (12)$$

For $n = 2$, as a special case, we get the classical AF [25] as

$$AF = AF_\gamma^{(2)} = 2 A_1^{-1} \frac{G_{2,6}^{6,1}\left[B \, \big| \, \begin{matrix} -1, \kappa_1 \\ \kappa_2 \end{matrix}\right]}{G_{2,6}^{6,1}\left[B \, \big| \, \begin{matrix} 0, \kappa_1 \\ \kappa_2 \end{matrix}\right]^2} - 1. \quad (13)$$

For the non-pointing errors case, when $\xi \to \infty$, it can be easily shown that the $n^{th}$-order AF in (12) converges to

$$AF_\gamma^{(n)} = n \, A_2^{1-n} \, G_{1,5}^{5,1}\left[B \, \bigg| \, \begin{matrix} 1-n \\ \kappa_3 \end{matrix}\right] G_{1,5}^{5,1}\left[B \, \bigg| \, \begin{matrix} 0 \\ \kappa_3 \end{matrix}\right]^{-n} - 1. \quad (14)$$

For $n = 2$, as a special case, we get the classical AF [25], for non-pointing errors case, as

$$AF = AF_\gamma^{(2)} = 2 A_2^{-1} \, G_{1,5}^{5,1}\left[B \, \bigg| \, \begin{matrix} -1 \\ \kappa_3 \end{matrix}\right] G_{1,5}^{5,1}\left[B \, \bigg| \, \begin{matrix} 0 \\ \kappa_3 \end{matrix}\right]^{-2} - 1. \quad (15)$$

TABLE I
BER PARAMETERS OF BINARY MODULATIONS

| Modulation | $p$ | $q$ |
| --- | --- | --- |
| Coherent Binary Frequency Shift Keying (CBFSK) | 0.5 | 0.5 |
| Coherent Binary Phase Shift Keying (CBPSK) | 0.5 | 1 |
| Non-Coherent Binary Frequency Shift Keying (NBFSK) | 1 | 0.5 |
| Differential Binary Phase Shift Keying (DBPSK) | 1 | 1 |

### B. Error Probability

*1) Average BER:* Substituting (4) into [26, Eq. (12)] and utilizing [21, Eq. (7.813.1)], we get the average BER $\overline{P}_b$ of a variety of binary modulations as

$$\overline{P}_b = \frac{1}{2} - \frac{A_1 \, q^p \, \Gamma(p)^{-1}}{2\,(q+1/\overline{\gamma}_1)^p} \, G_{2,6}^{6,1}\!\left[\frac{B}{q\,\overline{\gamma}_1 + 1}\,\bigg|\,\begin{matrix}1-p,\kappa_1\\\kappa_2\end{matrix}\right], \quad (16)$$

where the parameters $p$ and $q$ account for different modulation schemes. For an extensive list of modulation schemes represented by these parameters, one may look into [26]–[29] or refer to Table I. For the non-pointing errors case, when $\xi \to \infty$, the BER in (16) can be easily shown to converge to

$$\overline{P}_b = \frac{1}{2} - \frac{A_2 \, q^p \, \Gamma(p)^{-1}}{2\,(q+1/\overline{\gamma}_1)^p} \, G_{1,5}^{5,1}\!\left[\frac{B}{q\,\overline{\gamma}_1 + 1}\,\bigg|\,\begin{matrix}1-p\\\kappa_3\end{matrix}\right]. \quad (17)$$

*2) Average SER:* In [30], the conditional SER has been presented in a desirable form and utilized to obtain the average SER of M-AM, M-PSK, and M-QAM. For example, for M-PSK the average SER $\overline{P}_s$ over generalized fading channels is given by [30, Eq. (41)]. Similarly, for M-AM and M-QAM, the average SER $\overline{P}_s$ over generalized fading channels is given by [30, Eq. (45)] and [30, Eq. (48)] respectively. On substituting (8) into [30, Eq. (41)], [30, Eq. (45)], and [30, Eq. (48)], we can get the SER of M-PSK, M-AM, and M-QAM, as shown below

$$\overline{P}_s = \frac{M-1}{M} + \frac{A_1}{\pi} \int_0^{\frac{(M-1)\pi}{M}} \frac{\sin^2(\pi/M)/\sin^2\phi}{\left(1/\overline{\gamma}_1 - \sin^2(\pi/M)/\sin^2\phi\right)} \\ \times G_{2,6}^{6,1}\!\left[\frac{B}{1-\left(\sin^2(\pi/M)/\sin^2\phi\right)\overline{\gamma}_1}\,\bigg|\,\begin{matrix}0,\kappa_1\\\kappa_2\end{matrix}\right] d\phi, \quad (18)$$

$$\overline{P}_s = \frac{M-1}{M} + \frac{2 A_1 (M-1)}{M\pi} \\ \times \int_0^{\frac{\pi}{2}} \frac{3/\left((M^2-1)\sin^2\phi\right)}{\left(1/\overline{\gamma}_1 - 3/\left((M^2-1)\sin^2\phi\right)\right)} \\ \times G_{2,6}^{6,1}\!\left[\frac{B}{1-\left(3/\left((M^2-1)\sin^2\phi\right)\right)\overline{\gamma}_1}\,\bigg|\,\begin{matrix}0,\kappa_1\\\kappa_2\end{matrix}\right] d\phi, \quad (19)$$

and

$$\overline{P}_s = 2\left(1-1/\sqrt{M}\right) - \left(1-1/\sqrt{M}\right)^2 \\ + (4\,A_1/\pi)\left(1-1/\sqrt{M}\right) \\ \times \int_0^{\frac{\pi}{2}} \frac{3/\left(2\,(M-1)\sin^2\phi\right)}{\left(1/\overline{\gamma}_1 - 3/\left(2\,(M-1)\sin^2\phi\right)\right)} \\ \times G_{2,6}^{6,1}\!\left[\frac{B}{1-\left(3/\left(2\,(M-1)\sin^2\phi\right)\right)\overline{\gamma}_1}\,\bigg|\,\begin{matrix}0,\kappa_1\\\kappa_2\end{matrix}\right] \\ - (4\,A_1/\pi)\left(1-1/\sqrt{M}\right)^2 \\ \times \int_0^{\frac{\pi}{4}} \frac{3/\left(2\,(M-1)\sin^2\phi\right)}{\left(1/\overline{\gamma}_1 - 3/\left(2\,(M-1)\sin^2\phi\right)\right)} \\ \times G_{2,6}^{6,1}\!\left[\frac{B}{1-\left(3/\left(2\,(M-1)\sin^2\phi\right)\right)\overline{\gamma}_1}\,\bigg|\,\begin{matrix}0,\kappa_1\\\kappa_2\end{matrix}\right] d\phi, \quad (20)$$

respectively. The analytical SER performance expressions obtained in (18), (19), and (20) are exact and can be easily estimated accurately by utilizing the Gauss-Chebyshev Quadrature (GCQ) formula [31, Eq. (25.4.39)] that converges rapidly, requiring only few terms for an accurate result [11].

### C. Ergodic Capacity

The ergodic channel capacity $\overline{C}$ defined as $\overline{C} \triangleq \mathbb{E}\left[\log_2(1+\gamma)\right]$ can be expressed in terms of the CCDF of $\gamma$ as $\overline{C} = 1/\ln(2) \int_0^\infty (1+\gamma)^{-1} F_\gamma^c(\gamma)\,d\gamma$ [32, Eq. (15)]. Utilizing this equation by exploiting the identity [33, p. 152] $(1+az)^{-b} = \frac{1}{\Gamma(b)} G_{1,1}^{1,1}\!\left[az\,\big|\,\begin{matrix}1-b\\0\end{matrix}\right]$ in it and using the integral identity [26, Eq. (20)], the ergodic capacity can be expressed in terms of the extended generalized bivariate Meijer's G function (EGBMGF) (see [26] and references therein) as

$$\overline{C} = \frac{A_1 \overline{\gamma}_1}{\ln(2)} \, G_{1,0:1,1:1,6}^{1,0:1,1:6,0}\!\left[1\,\bigg|\,\begin{matrix}0\\0\end{matrix}\,\bigg|\,\begin{matrix}\kappa_1\\\kappa_2\end{matrix}\,\bigg|\,\overline{\gamma}_1, B\right]. \quad (21)$$

For the non-pointing errors case, when $\xi \to \infty$, the ergodic capacity in (21) can be easily shown to converge to

$$\overline{C} = \frac{A_2 \overline{\gamma}_1}{\ln(2)} \, G_{1,0:1,1:0,5}^{1,0:1,1:5,0}\!\left[1\,\bigg|\,\begin{matrix}0\\0\end{matrix}\,\bigg|\,\kappa_3\,\bigg|\,\overline{\gamma}_1, B\right]. \quad (22)$$

The expression in (21) and (22) can be easily and efficiently evaluated by utilizing the MATHEMATICA® implementation of the EGBMGF given in [26, Table II].

## V. RESULTS AND DISCUSSION

The average BER performance of different digital binary modulation schemes are presented in Fig. 2 based on the values of $p$ and $q$ as presented in Table I. We can observe from Fig. 2 that the simulation results provide a perfect match to the analytical results obtained in this work.

It can be seen from Fig. 2 that, as expected, CBPSK outperforms NBFSK. Also, the effect of pointing error can be observed in Fig. 2 i.e. as the effect of pointing error (as the value of $\xi$ increases, the effect of pointing error decreases) increases, the BER deteriorates and vice versa. It can be shown that as the atmospheric turbulence conditions get severe i.e.

## VI. CONCLUDING REMARKS

We derived novel exact closed-form expressions for the CDF, the PDF, the MGF, and the moments of an asymmetric dual-hop relay transmission system composed of both RF and FSO environments with pointing errors in terms of Meijer's G functions. Further, we derived analytical expressions for various performance metrics of an asymmetric dual-hop RF/FSO relay transmission system with pointing errors including the higher-order AF, error rate of a variety of modulation schemes, and the ergodic capacity in terms of Meijer's G functions. In addition, this work presents simulation examples to validate and illustrate the mathematical formulation developed in this work and to show the effect of the atmospheric turbulence and pointing error conditions severity and unbalance on the system performance.

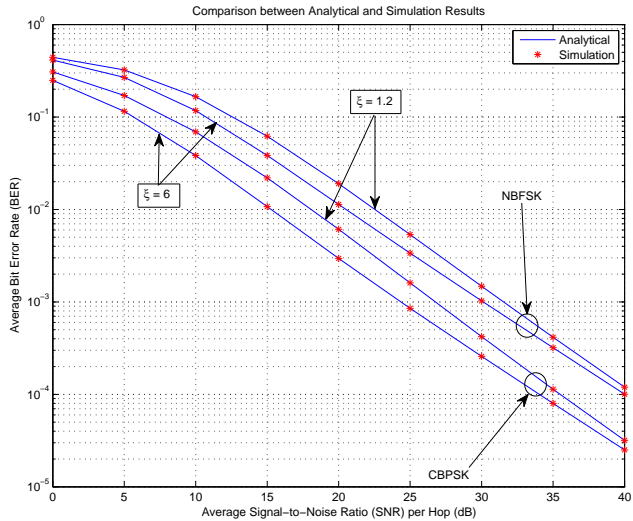

Fig. 2. Average BER of different binary modulation schemes showing impact of pointing errors (varying $\xi$) with fading parameters $\alpha = 2.1$, $\beta = 3.5$, and $C = 0.6$.

as the values of $\alpha$ and $\beta$ start dropping, the BER starts deteriorating and vice versa. Similar results for any other binary modulations schemes and any other values of $\alpha$'s, $\beta$'s, $c$'s, and $\xi$'s can be observed.

Similarly, in Fig. 3, as the atmospheric turbulence conditions get severe, the ergodic capacity starts decreasing (i.e. the higher the values of $\alpha$ and $\beta$, the higher will be the ergodic capacity). Also, the effect of pointing error can be observed in Fig. 3. Note that as the value of $\xi$ increases (i.e. the effect of pointing error decreases) the ergodic capacity decreases.

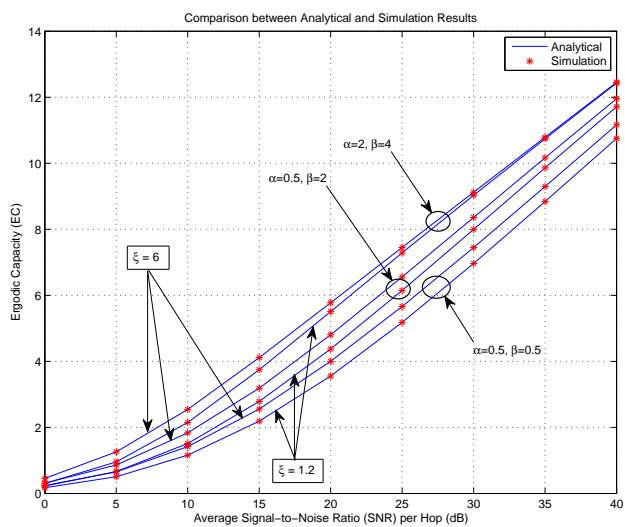

Fig. 3. Effect of pointing errors (varying $\xi$) on the ergodic capacity with varying fading parameters $\alpha$'s and $\beta$'s, and $C = 0.6$.

## REFERENCES

[1] E. Lee, J. Park, D. Han, and G. Yoon, "Performance analysis of the asymmetric dual-hop relay transmission with mixed RF/FSO links," *IEEE Photonics Technology Letters*, vol. 23, no. 21, pp. 1642–1644, Nov. 2011.

[2] L. C. Andrews, R. L. Phillips, and C. Y. Hopen, *Laser Beam Scintillation with Applications*. Bellingham, WA: SPIE, 2001.

[3] K. P. Peppas and C. K. Datsikas, "Average symbol error probability of general-order rectangular quadrature amplitude modulation of optical wireless communication systems over atmospheric turbulence channels," *IEEE/OSA Journal of Optical Communications and Networking*, vol. 2, no. 2, pp. 102–110, Feb. 2010.

[4] W. O. Popoola and Z. Ghassemlooy, "BPSK subcarrier intensity modulated free-space optical communications in atmospheric turbulence," *IEEE/OSA Journal of Lightwave Technology*, vol. 27, no. 8, pp. 967–973, Apr. 2009.

[5] J. Park, E. Lee, and G. Yoon, "Average bit error rate of the Alamouti scheme in Gamma-Gamma fading channels," *IEEE Photonics Technology Letters*, vol. 23, no. 4, pp. 269–271, Feb. 2011.

[6] M. Safari and M. Uysal, "Relay-assisted free-space optical communication," *IEEE Transactions on Wireless Communications*, vol. 7, no. 12, pp. 5441–5449, Dec. 2008.

[7] H. G. Sandalidis, T. A. Tsiftsis, G. K. Karagiannidis, and M. Uysal, "BER performance of FSO links over strong atmospheric turbulence channels with pointing errors," *IEEE Communications Letters*, vol. 12, no. 1, pp. 44–46, Jan. 2008.

[8] H. G. Sandalidis, T. A. Tsiftsis, and G. K. Karagiannidis, "Optical wireless communications with heterodyne detection over turbulence channels with pointing errors," *Journal of Lightwave Technology*, vol. 27, no. 20, pp. 4440–4445, Oct. 2009.

[9] W. Gappmair, "Further results on the capacity of free-space optical channels in turbulent atmosphere," *IET Communications*, vol. 5, no. 9, pp. 1262–1267, Jun. 2011.

[10] M. O. Hasna and M.-S. Alouini, "A performance study of dual-hop transmissions with fixed gain relays," *IEEE Transactions on Wireless Communications*, vol. 3, no. 6, pp. 1963–1968, Nov. 2004.

[11] F. Yilmaz, O. Kucur, and M.-S. Alouini, "A novel framework on exact average symbol error probabilities of multihop transmission over amplify-and-forward relay fading channels," in *Proceedings of $7^{th}$ International Symposium on Wireless Communication Systems (ISWCS' 2010)*, York, U.K., Nov. 2010, pp. 546–550.

[12] Y. Zhu, Y. Xin, and P.-Y. Kam, "Outage probability of Rician fading relay channels," *IEEE Transactions on Vehicular Technology*, vol. 57, no. 4, pp. 2648–2652, Jul. 2008.

[13] S. N. Datta, S. Chakrabarti, and R. Roy, "Error analysis of non coherent FSK with variable gain relaying in dual-hop Nakagami-$m$ relay fading channel," in *Proceedings of 2010 International Conference on Signal Processing and Communications (SPCOM' 2010)*, Bangalore, India, Jul. 2010, pp. 1–5.

[14] H. A. Suraweera, R. H. Y. Louie, Y. Li, G. K. Karagiannidis, and B. Vucetic, "Two hop amplify-and-forward transmission in mixed Rayleigh and Rician fading channels," *IEEE Communication Letters*, vol. 13, no. 4, pp. 227–229, Apr. 2009.

[15] A. K. Gurung, F. S. Al-Qahtani, Z. M. Hussain, and H. Alnuweiri, "Performance analysis of amplify-forward relay in mixed Nakagami-$m$ and Rician fading channels," in *Proceedings of 2010 International Conference on Advanced Technologies for Communications (ATC' 2010)*, Ho Chi Minh City, Vietnam, Oct. 2010, pp. 321–326.

[16] F. Yilmaz and M.-S. Alouini, "Product of the powers of generalized Nakagami-$m$ variates and performance of cascaded fading channels," in *Proceedings of IEEE Global Telecommunications Conference, 2009. (GLOBECOM 2009)*, Honolulu, Hawaii, US, Nov.-Dec. 2009, pp. 1–8.

[17] C. K. Datsikas, K. P. Peppas, N. C. Sagias, and G. S. Tombras, "Serial free-space optical relaying communications over Gamma-Gamma atmospheric turbulence channels," *IEEE/OSA Journal of Optical Communications and Networking*, vol. 2, no. 8, pp. 576–586, Aug. 2010.

[18] N. Saquib, M. S. R. Sakib, A. Saha, and M. Hussain, "Free space optical connectivity for last mile solution in Bangladesh," in *Proceedings of $2^{nd}$ International Conference on Education Technology and Computer (ICETC' 10)*, Shanghai, China, Jun. 2010, pp. 484–487.

[19] M. K. Simon and M.-S. Alouini, *Digital Communication over Fading Channels*, 2nd ed. Hoboken, New Jersey, USA: IEEE: John Wiley & Sons, Inc., 2005.

[20] M. D. Springer, *The Algebra of Random Variables*. New York: Wiley, Apr. 1979.

[21] I. S. Gradshteyn and I. M. Ryzhik, *Table of Integrals, Series and Products*. New York: Academic Press, 2000.

[22] I. Wolfram Research, *Mathematica Edition: Version 8.0*. Champaign, Illinois: Wolfram Research, Inc., 2010.

[23] V. S. Adamchik and O. I. Marichev, "The algorithm for calculating integrals of hypergeometric type functions and its realization in reduce system," in *Proceedings of International Symposium on Symbolic and Algebraic Computation (ISSAC' 90)*, New York, USA, 1990, pp. 212–224.

[24] F. Yilmaz and M.-S. Alouini, "Novel asymptotic results on the high-order statistics of the channel capacity over generalized fading channels," in *Proceedings of IEEE $13^{th}$ International Workshop on Signal Processing Advances in Wireless Communications (SPAWC' 2012)*, Cesme, Turkey, Jun. 2012, pp. 389–393.

[25] U. Charash, "Reception through Nakagami fading multipath channels with random delays," *IEEE Transactions on Communications*, vol. 27, no. 4, pp. 657–670, Apr. 1979.

[26] I. S. Ansari, S. Al-Ahmadi, F. Yilmaz, M.-S. Alouini, and H. Yanikomeroglu, "A new formula for the BER of binary modulations with dual-branch selection over generalized-$K$ composite fading channels," *IEEE Transactions on Communications*, vol. 59, no. 10, pp. 2654–2658, Oct. 2011.

[27] N. C. Sagias, D. A. Zogas, and G. K. Kariaginnidis, "Selection diversity receivers over nonidentical Weibull fading channels," *IEEE Transactions on Vehicular Technology*, vol. 54, no. 6, pp. 2146–2151, Nov. 2005.

[28] A. H. Wojnar, "Unknown bounds on performance in Nakagami channels," *IEEE Transactions on Communications*, vol. 34, no. 1, pp. 22–24, Jan. 1986.

[29] I. S. Ansari, F. Yilmaz, and M.-S. Alouini, "On the sum of Gamma random variates with application to the performance of maximal ratio combining over Nakagami-$m$ fading channels," in *Proceedings of IEEE $13^{th}$ International Workshop on Signal Processing Advances in Wireless Communications (SPAWC' 2012)*, Cesme, Turkey, Jun. 2012, pp. 394–398.

[30] M.-S. Alouini and A. J. Goldsmith, "A unified approach for calculating error rates of linearly modulated signals over generalized fading channels," *IEEE Transactions on Communications*, vol. 47, no. 9, pp. 1324–1334, Sep. 1999.

[31] M. Abramowitz and I. A. Stegun, *Handbook of Mathematical Functions*, 10th ed. New York: Dover, Dec. 1972.

[32] A. Annamalai, R. C. Palat, and J. Matyjas, "Estimating ergodic capacity of cooperative analog relaying under different adaptive source transmission techniques," in *Proceedings of 2010 IEEE Sarnoff Symposium*, Princeton, NJ, Apr. 2010, pp. 1–5.

[33] A. M. Mathai and R. K. Saxena, *The H-Function with Applications in Statistics and Other Disciplines*. New York: Wiley Eastern, 1978.